\documentclass[aps,prl,preprint,groupedaddress]{revtex4}

\usepackage{graphicx}

\begin{document}

\title{Hot electron attenuation of direct and scattered carriers across an epitaxial Schottky interface} 

\author{S. Parui$^1$, P. S. Klandermans$^1$, S. Venkatesan$^2$, C. Scheu$^2$, and T. Banerjee$^1$}

 \email{T.Banerjee@rug.nl}
\affiliation{$^1$Physics of Nanodevices, Zernike Institute for Advanced Materials, University of Groningen, Nijenborgh 4, 9747 AG Groningen, The Netherlands}
\affiliation{$^2$Department of Chemistry and Center for NanoScience, Ludwig-Maximilians-Universit$\ddot{a}$t M$\ddot{u}$nchen, Butenandstr 5-13(E), 81377 M$\ddot{u}$nchen, Germany}
 
\date{\today}

\begin{abstract}
Hot electron transport of direct and scattered carriers across an epitaxial NiSi$_2$/n-Si(111) interface, for different NiSi$_2$ thickness, is studied using Ballistic Electron Emission Microscopy (BEEM). We find the BEEM transmission for the scattered hot electrons in NiSi$_2$ to be significantly lower than that for the direct hot electrons, for all thicknesses. Interestingly, the attenuation length of the scattered hot electrons is found to be twice larger than that of the direct hot electrons. The lower BEEM transmission for the scattered hot electrons is due to inelastic scattering of the injected hot holes while the larger attenuation length of the scattered hot electrons is a consequence of the differences in the energy distribution of the injected and scattered hot electrons and the increasing attenuation length, at lower energies, of the direct hot electrons in NiSi$_2$.

\vspace{1pc}
\vspace{1pc}  
\end{abstract}

\maketitle
\section{\label{sec:level1}Introduction}
Hot electron transport has been widely employed in studies related to the relaxation and dynamics of excited electrons in different physical and chemical processes ranging from electronic transport, optical and two-photon photoemission experiments, surface chemistry, strong correlations in transition-metal oxides etc.\cite{Kaiser-Bell,twophoton,Rippard,TamalikaFe,Yi,Parui,Nienhaus,RanaSciRep}. Hot electron scattering has also been studied using $ab$-$initio$ techniques, by combining first principles approach based on density functional theory with many-body perturbation theory yielding insights into the role of electron-phonon scattering, contribution of the $d$ electrons to screening as well as scattering and overestimation of the scattering rates using the free-electron model \cite{zhukovprb,ClaudiaPRB, Petek}. In spite of these studies, very little is known about the scattering processes and transport of secondary electron-hole ($\textit{e-h}$) pairs that are created in Auger-like scattering processes in such experiments. Transport studies using Ballistic Electron Emission Microscopy (BEEM) have a particular advantage in this regard as such processes can be easily studied using the same device structure as that employed to study direct hot electron scattering. In this work, we demonstrate hot electron scattering and attenuation in a model epitaxial Schottky interface of NiSi$_2$/n-Si(111) using the different modes in BEEM \cite{Kaiser-Bell,Bell-reverse, Haq}. We do this by changing the injection bias polarity of the scanning tunneling microscope (STM) tip that is used in the BEEM, such that we study hot electron attenuation of both the direct and scattered electrons by injecting hot electrons and hot holes respectively. 
Using such an epitaxial interface, where the transmission probability has been demonstrated to be large for hot electrons \cite{SubirAPL}, one can study hot electron scattering processes that are predominantly determined by inelastic $e-e$ scattering and less sensitive to the momentum scattering in the epitaxial NiSi$_2$ layers or at the epitaxial Schottky interface. We find that the BEEM transmission for the scattered hot electrons is lower than the BEEM transmission with the direct hot electrons. Furthermore, what is interesting is that the hot electron attenuation length, $\lambda$, associated with the scattered hot electrons ($\lambda_{eff}$) in NiSi$_2$ is twice larger than that with direct hot electrons. The lower BEEM transmission for the scattered hot electrons is due to inelastic scattering of the injected hot holes, that removes electrons with lower energy from being collected, while the longer attenuation length of the scattered hot electrons is a cumulative effect of the differences in the energy distribution of the injected and scattered hot electrons and the increasing attenuation length, at lower energies, of the direct hot electrons in NiSi$_2$ as measured using BEEM.\\
\section{\label{sec:level1}Experimental Technique}
\indent BEEM uses the tip of a scanning tunneling microscope (STM) to inject hot electrons (energy few eV above the Fermi level) through a vacuum tunnel barrier into a thin metal (M) layer forming a Schottky contact on a n-type semiconductor (S) as shown in Fig. 1. A fraction of the electrons injected in the NiSi$_2$ film is collected in the semiconductor as the collector current ($I_B$), if they satisfy the energy and momentum criteria at the M/S Schottky interface (Fig. 2(a)). The epitaxial Schottky interface of NiSi$_2$/n-Si(111) is shown in Fig. 1(b), the inset of which shows a High Resolution Transmission Electron Microscopy image of a Type-A NiSi$_2$ on n-Si(111). By placing the STM tip at different locations of the device, the local Schottky barrier height can be extracted using the Bell-Kaiser (B-K) model \cite{Kaiser-Bell} which states that $I_{B}  {\propto} (V_T-\phi_B)^2$, where $V_T$ is the applied tip voltage and $\phi_B$ is the Schottky barrier height at the M/S interface. This mode of BEEM has been successfully applied to study transport across various M/S interfaces and probing the spatial homogeneity of transport across such interfaces \cite{Kaiser-Bell, Parui, Palm, Ludeke, Bobisch, SubirAPL}. BEEM can also be operated in a reverse mode, known as the reverse BEEM (R-BEEM), that is realized by applying a positive tip bias \cite{Bell-reverse} as shown in Fig. 2(b). In R-BEEM, hot holes that are injected in the NiSi$_2$ layer scatters with the electron gas close to $E_F$ and creates secondary electrons by electron-hole ($\textit{e-h}$) pair generation, similar to the Auger scattering process. These scattered electrons are then collected in the n-Si(111) as a Reverse BEEM current ($I_{RB}$) which, near threshold, follows a power four dependence with the injected bias \cite{Bell-reverse} i.e. $I_{RB}  {\propto} (V_T-\phi_B)^4$. The energy distribution of the injected tunnel electrons is represented in Fig. 2(c) in the direct BEEM for a negative tip bias i.e $eV_T$. For the revere BEEM, the distribution of the injected holes at a positive tip bias of -$eV_T$ is shown in Fig. 2(d) together with the scattered electron distribution. The distribution of the injected electrons and the injected holes correspond to the direct tunneling probability between the STM tip and the metal base for both modes, whereas the distribution of the excited electrons in the reverse BEEM arises due to the inelastic scattering at the metal base by the injected hot holes. The distribution of the injected hot electrons are maximum at the $E_F$ of the STM tip for the direct BEEM whereas it is peaked at the $E_F$ of the metal base for the R-BEEM. For a M/S interface with a Schottky barrier height (SBH) of $\phi_B$ as shown, this suggests that a large fraction of the more energetic electrons can be collected in direct BEEM whereas in R-BEEM only a small fraction of the scattered electrons can be collected close to $\phi_B$ that slowly increases with energy. Hot electron attenuation length $(\lambda)$ for both the direct and scattered carriers can be measured in such metal layers across different semiconductor interfaces by considering the exponential dependence of the BEEM and R-BEEM transmissions with base layer thicknesses $(t)$ as $I_{B}(t, E){\propto} exp\left[-t/\lambda(E)\right]$ where $E$ is the energy of the hot carriers. From R-BEEM studies, $\lambda_{eff}$ can be extracted which depends not only on the attenuation length of the injected hot holes  but also on the attenuation length of the scattered electrons \cite{Bell-reverse, Niedermann}. It is non-trivial to decouple the exact contribution of the different scattering processes in the extraction of $\lambda_{eff}$ for the scattered carriers.\\
\indent In this work, we investigate the thickness dependence of BEEM transmission for both the direct and scattered electrons in NiSi$_2$ grown epitaxially on n-Si(111). Such an epitaxial Schottky interface (lattice mismatch of 0.46$\%$) with demonstrated large transmissions for hot electrons \cite{SubirAPL} are ideally suited for the study of inelastic scattering of the injected carriers, as, at such epitaxial films and interfaces the contribution of elastic scattering to hot electron attenuation is expected to be minimal. This will also enhance the propagation and collection of those electrons with momentum parallel to the M/S interface, i.e parallel momentum $(k_{||})$ is conserved, for both the direct and scattered electrons at such interfaces. The energy dependence of the BEEM transmission also enables us to extract the attenuation length for both the direct and scattered electrons from the exponential decay of the collector current with NiSi$_2$ thickness.\\
\section{\label{sec:level1}Experimental Details}
\indent For this study, devices are fabricated on a patterned n-Si(111) substrate as described in Ref. \cite{SubirNi}. Initially Ni layers of varying thicknesses are deposited on chemically terminated Si(111) substrates \cite{SubirNi, Dumas}. Epitaxial NiSi$_2$ films are formed due to thermal annealing of the deposited Ni layer, according to the well established protocol \cite{SubirAPL, Tung}. Thereafter, a 4 nm thick Au capping layer is deposited at room temperature. The devices are then transferred $ex$ $situ$ to the BEEM set up. Electrical characterization of the diodes are performed by standard current-voltage (I-V) measurements. BEEM measurements are performed at LT (100 K) by a modified commercial STM from RHK. The sample top metal surface is grounded by using Au contact pad and a mechanically cut PtIr STM tip is used to inject the hot electrons for direct BEEM and hot holes for R-BEEM. A large area ohmic contact to the n-Si(111) substrate is used for hot electron collection in both cases.\\
\section{\label{sec:level1}Results and Discussions}
\indent Hot electron transmission for both the direct and reverse mode in BEEM are plotted in Fig. 3 as a function of tip bias, $V_T$, at a constant injection current, $I_T$, for different thicknesses of NiSi$_2$. Each spectrum is an average of $\sim$100 individual spectra taken at several different locations on the same device. Two observations are central to Fig. 3: i) with increasing thickness of NiSi$_2$, the BEEM and R-BEEM transmissions both decreases. For both cases, the transmission increases above a certain threshold that corresponds to the SBH at the NiSi$_2$/n-Si interface ($\phi_B$) and ii) the energy dependence of the reverse BEEM transmission is less pronounced for all NiSi$_2$ thickness as compared to the direct BEEM transmission which shows a marked dependence on energy for all thicknesses. The R-BEEM transmissions in Fig. 3 have been multiplied by 20. The spectral shape for the reverse BEEM is also different than the direct BEEM as can be clearly observed by normalizing both the plots at 1.8 V (for the 4 nm NiSi$_2$ film), as shown in the inset of Fig. 3. This is easily understood from the B-K model which states that close to the threshold, the direct BEEM and R-BEEM transmission varies as power 2 and 4 respectively, above $\phi_B$, and further indicates the different energy dependence of scattering for the two processes.\\
\indent The ratio of the energy dependence of $I_{B}$ to I$_{RB}$, is plotted in Fig. 4 and represents the efficiency of collection of the scattered electrons created by electron-hole ($\textit{e-h}$) pair generation in R-BEEM. An interesting trend is found in this ratio viz. the ratio increases sharply with decreasing tip bias and for all film thicknesses. For example, at 1 V tip bias, this ratio is 80 for the 24 nm NiSi$_2$ film decreasing to 10 at 1.8 V, while it is 280 at 1 V for the 4 nm NiSi$_2$ film that decreases to 20 at 1.8 V.\\
\indent Besides the small fraction of hot electrons that may reach the epitaxial M/S interface without scattering, there can also be contribution to $I_{B}$ from the inelastic scattering of the injected hot electrons. During such an inelastic scattering event, a hot electron can maximally lose 50$\%$ of its energy and from the energy distribution of the injected hot electrons, as shown in Fig. 2 (c), this clearly signifies that the probability of a scattered electron at lower energies to surmount the SBH is small giving rise to a decreased BEEM transmission at lower energies. For the injected hole distribution as in the R-BEEM (shown in Fig. 2(d)), only those secondary electrons created during the electron-hole pair generation are collected that originates from the tail of the distribution and are also few in number, thus leading to a much reduced R-BEEM transmission. What is interesting here is that the R-BEEM transmission is less sensitive to the NiSi$_2$ thickness. This is because an increasing film thickness favors inelastic scattering events which creates a larger number of scattered hot electrons that can be collected at the M/S interface. As the R-BEEM transmission, near threshold, includes an extra $(V_T-\phi_B)^2$ factor dependence with respect to the direct BEEM, a plot of (I$_{RB}$/I$_B$)$^{1/2}$ with tip bias is expected to be linear as is shown in the inset of Fig. 4.\\
\indent By plotting the direct BEEM transmission versus the film thickness, the hot electron attenuation length in NiSi$_2$ is extracted. Figure 5(a) shows the BEEM transmissions at $V_T$= -1.6 V and -1.2 V with varying NiSi$_2$ thicknesses. Solid lines are fits to the exponential decay and the extracted $\lambda$'s are 12.6 $\pm$ 1.2 nm and 14.2 $\pm$ 1.4 nm for the respective tip biases. The attenuation lengths are extracted similarly at various other tip bias and shown in Fig. 5(c). Similarly, from the R-BEEM transmission, the effective attenuation length for the scattered carriers are extracted for tip biases from 1.2 V to 1.8 V. The extracted $\lambda_{eff}$'s are 38.0 $\pm$ 3.2 nm at 1.2 V and 26.0 $\pm$ 3.2 nm at 1.6 V respectively as shown in Fig. 5(b). The energy dependence of the attenuation lengths are given in Fig. 5(d). Our observations of $\lambda_{eff}>\lambda$ is also consistent with a previous report on PtSi/Si \cite{Niedermann}. The reduced signal to noise ratio, close to $\phi_B$, introduces a large error in the extraction of the attenuation lengths for both the direct and scattered carriers in NiSi$_2$ and is thus not performed.\\
\indent From Fig. 5 (c) and (d), we see that $\lambda_{eff}$ is $\sim$ twice larger than $\lambda$ and has a different energy dependence. For the direct electrons, $\lambda$ is almost constant at higher energies whereas it increases with decreasing energy. For the scattered electrons, $\lambda_{eff}$ sharply increases with decreasing energy and becomes a constant only at the highest energies measured. This trend with direct electrons reflect a cumulative effect of the conservation of parallel momentum of the hot electrons close to $\phi_B$ at such epitaxial M/S interfaces, as well as the availability of the density of states in NiSi$_2$ \cite{BisiDOS}, at energies that are relevant for our studies, as explained next. For such epitaxial films and M/S interfaces and for energies close to $\phi_B$, the propagating hot electrons reaching the Schottky interface are considerably forward focused due to minimal elastic scattering and can be easily collected at the n-Si(111) semiconductor leading to an increasing $\lambda$ at these energies. At higher energies (i.e $E_F$ + 2 eV), the density of states in NiSi$_2$ is almost constant as is reflected in the orbital character of the states involved in NiSi$_2$ \cite{BisiDOS} resulting in a constant $\lambda$, for the direct electrons, at these energies. An interesting consequence of the enhancement of the attenuation length for the direct hot electrons in NiSi$_2$, at low energies, can be seen in Figs. 3 and 5 (d). In R-BEEM the transmission and collection at the epitaxial M/S Schottky interface is that of the scattered hot electrons that are created during inelastic scattering of the injected hot holes. These scattered hot electrons have lower energies but as shown in Fig. 5(c) are those which have a larger attenuation length and thus contributes to I$_{RB}$ and leads to an increase in $\lambda_{eff}$. This also leads to a less sensitivity of the R-BEEM transmission with increasing thickness as is shown in Fig. 3. Further, we see that the density of states below $E_F$ \cite{BisiDOS} (i.e the injected hot hole distribution) sharply rises with energy due to the contribution of the $d$ electrons and that is reflected in the energy dependence of $\lambda_{eff}$. All the above factors thus explain the larger $\lambda_{eff}$ as compared to $\lambda$ in NiSi$_2$ and their associated energy dependence. \\
\section{\label{sec:level1}Summary}
\indent In conclusion, we have used an epitaxial model system of NiSi$_2$/Si(111), with a large transmission probability for hot electrons, to investigate hot electron transport and attenuation of the direct and scattered carriers using BEEM and R-BEEM respectively. We show that the R-BEEM transmission is significantly lower than that of the direct BEEM while their energy dependence exhibits features that reflects the energy distribution of the injected and scattered electrons, the role of conservation of parallel momentum in such epitaxial system close to $\phi_B$ and the density of states in NiSi$_2$. All these leads to an attenuation length for hot electrons that is almost twice larger for the scattered electrons than for the direct electrons in NiSi$_2$. Our results will not only enhance the understanding of the role of scattered carriers in different physical and chemical phenomena but forms an important model system for theoretical analysis of the scattering rates of the excited carriers. These results are also relevant for designing devices, as such epitaxial interfaces have a high thermal stability and are commonly used as contacts in complementary-metal-oxide-semiconductor (CMOS) technology.\\
{\bf ACKNOWLEDGEMENTS}\\
\indent We thank  K. G. Rana and A. M. Kamerbeek for scientific discussions. We are grateful to P. Rudolf for use of the Molecular beam epitaxy deposition system. We acknowledge the financial support from the Netherlands Organization for Scientific Research NWO-VIDI program, the Zernike Institute for Advanced Materials and NanoNed program coordinated by the Dutch Ministry of Economic Affairs.\\

\clearpage
\begin{figure}
\includegraphics[scale=0.36]{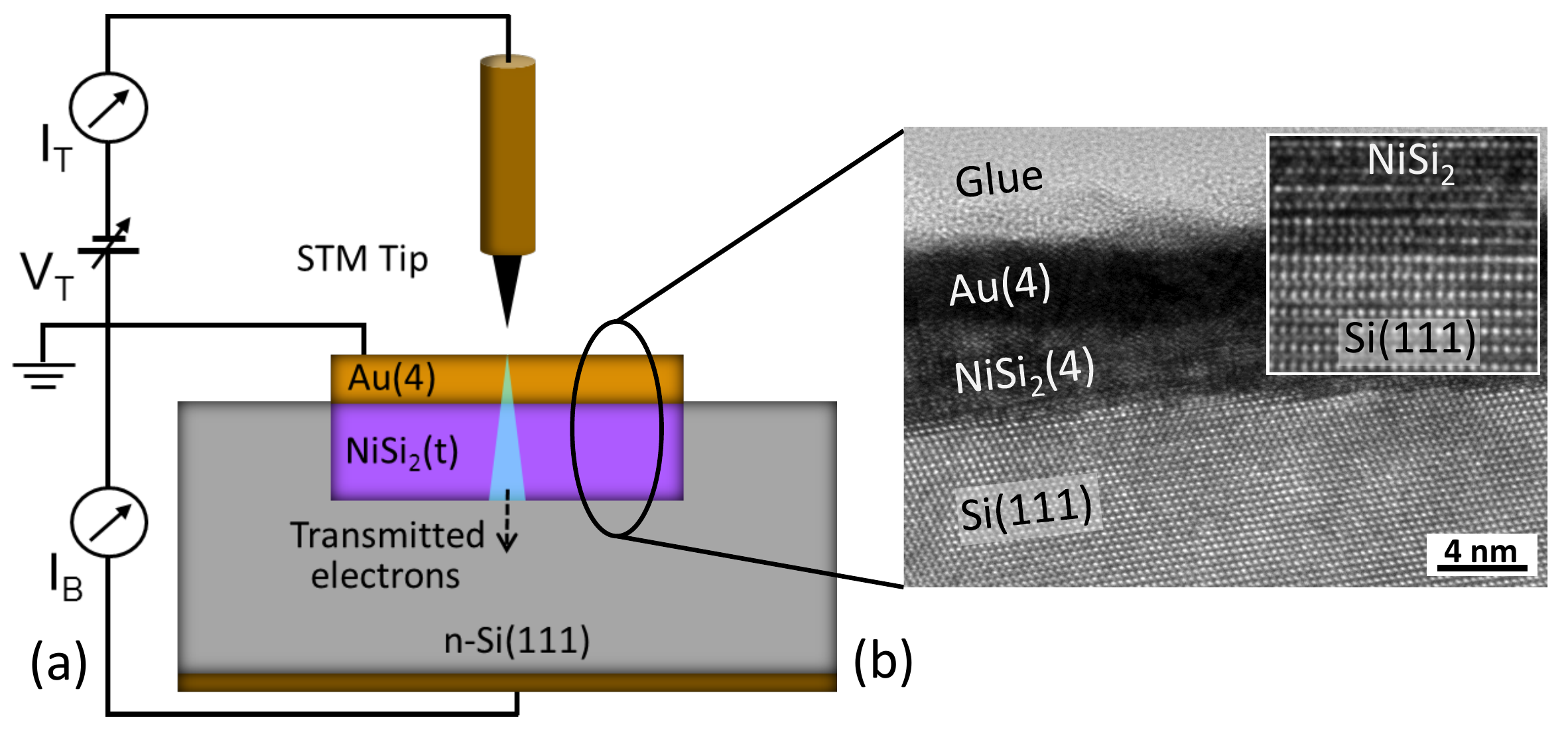}
\caption{\label{fig:TEM} (a) Device schematic of Ballistic Electron Emission Microscopy. (b) A Transmission Electron Microscopy (TEM) image of an NiSi$_2$/n-Si(111) device with a thin Au capping layer, viewed in a $<$1\={1}0$>$ cross section. The inset shows a high resolution TEM image of a Type-A NiSi$_2$ interface on n-Si(111). }
\end{figure}

\begin{figure}
\includegraphics[scale=0.46]{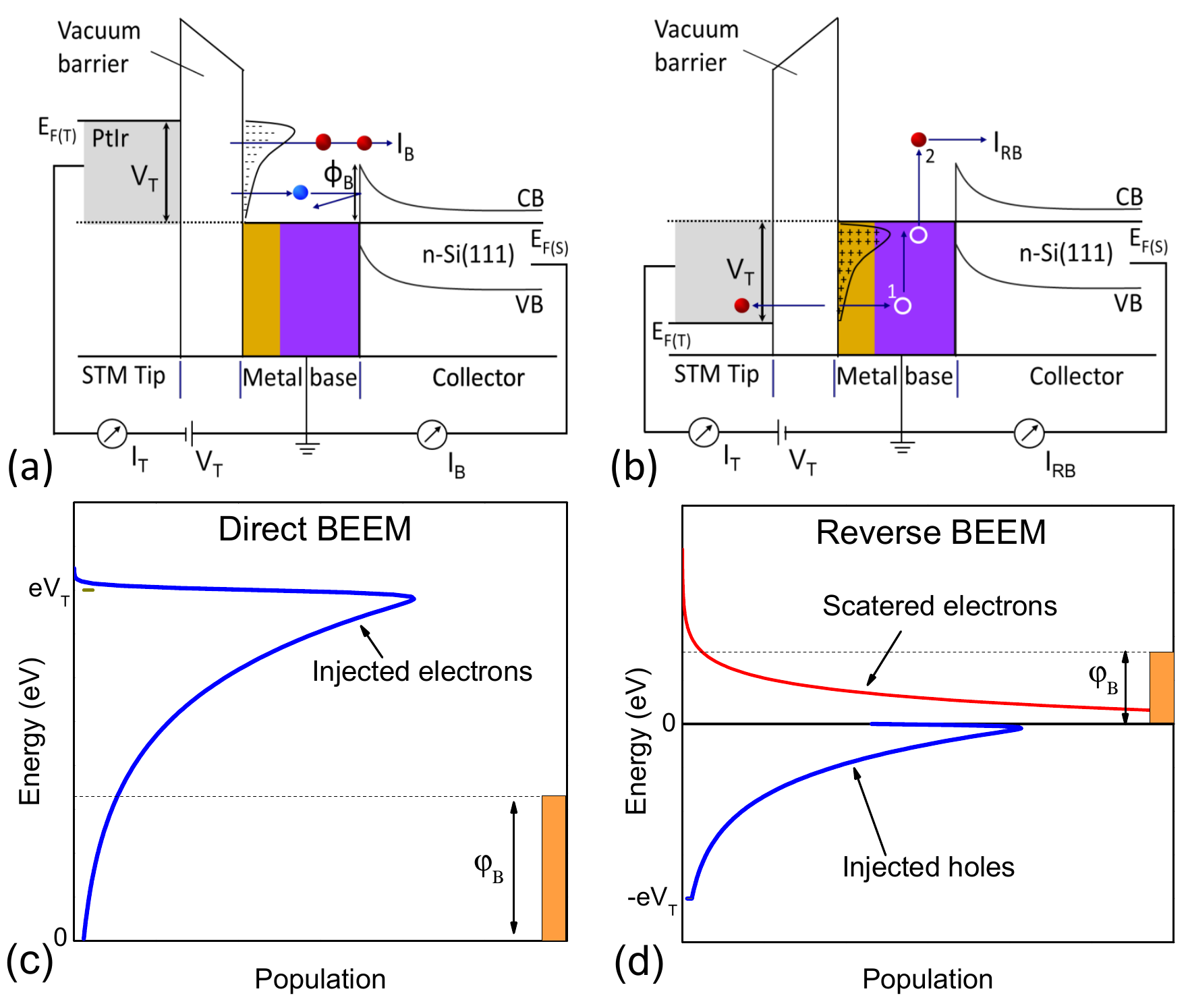}
\caption{\label{fig:TEM} (a) Energy schematic of direct BEEM: Hot electrons are injected and a fraction of them are collected. (b) Energy schematic of reverse BEEM: The injected hot holes creates secondary (scattered) electrons by $\textit{e-h}$ pairs which are then collected. (c) Schematic distribution of the injected hot electrons for the direct BEEM. Electrons with energies above the Schottky barrier height, $\phi_B$, will have a larger probability to be collected. (d) The distribution of the injected hot holes in reverse BEEM along with the excited electron distribution after $\textit{e-h}$ scattering.}
\end{figure}

\begin{figure}
\includegraphics[scale=0.39]{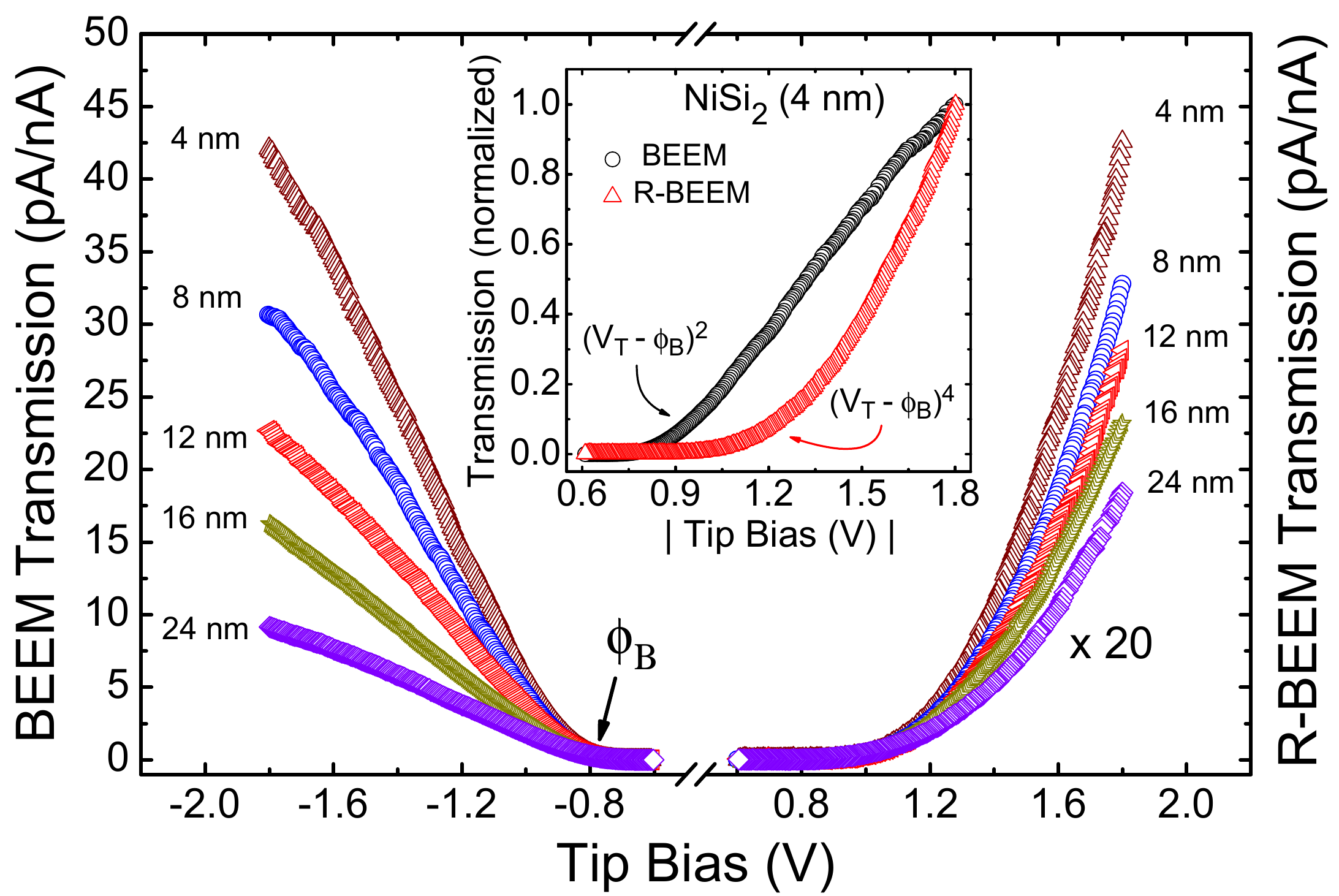}
\caption{\label{fig:BEEM and RBEEM} Representative BEEM and R-BEEM transmission vs tip voltage for Au(4 nm)/NiSi$_2$(t)/n-Si(111) devices, with varying NiSi$_2$ thickness. $\phi_B$ represents the Schottky barrier at the NiSi$_2$/n-Si(111) interface. Inset shows the energy dependence of the direct BEEM and reverse BEEM spectra for the 4 nm NiSi$_2$ film normalized at 1.8 V. }
\end{figure}

\begin{figure}
\includegraphics[scale=0.29]{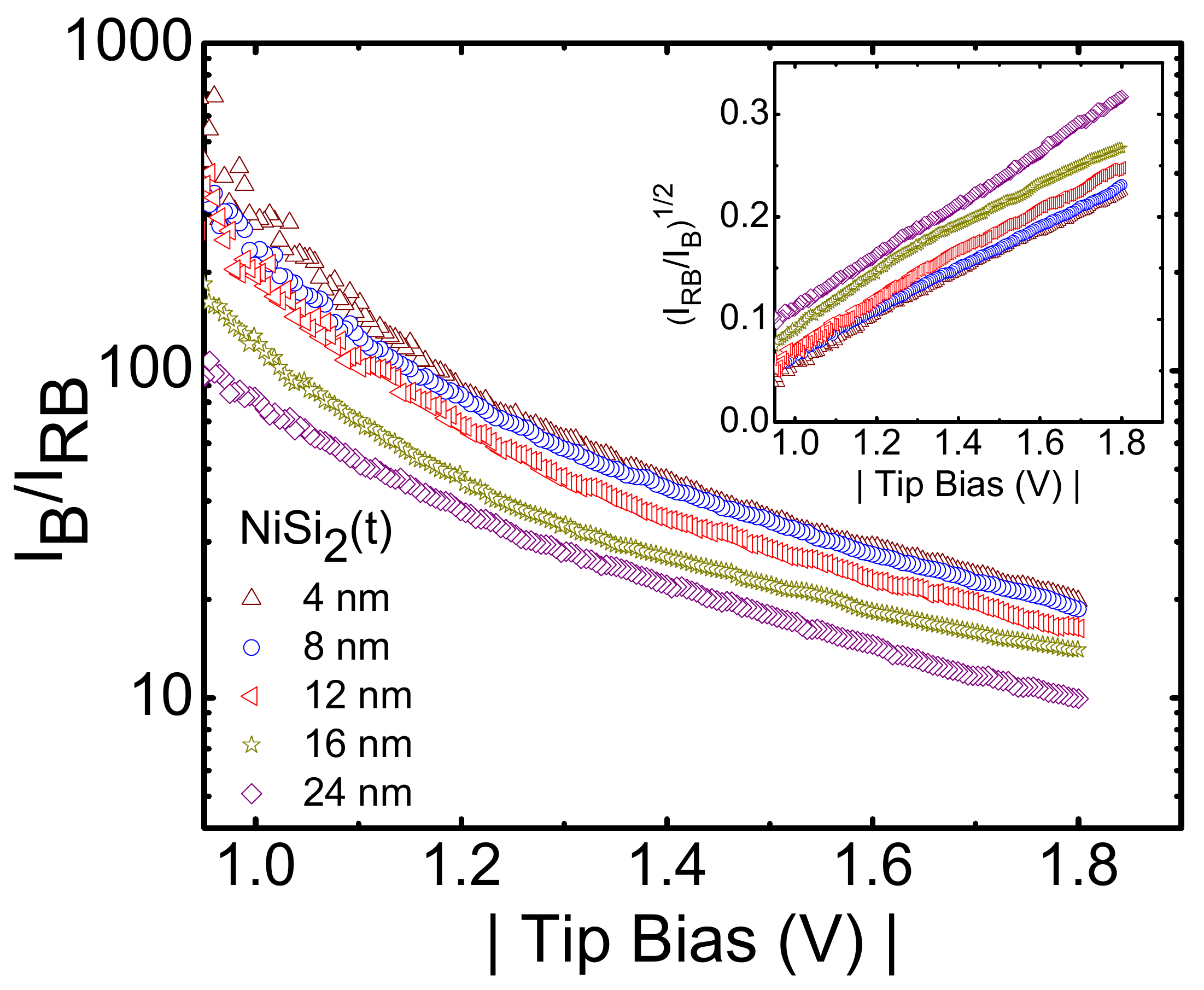}
\caption{\label{fig:Ratio} (a) Ratio of I$_B$/I$_{RB}$ with respect to the applied bias for all NiSi$_2$ thickness, manifesting the creation efficiency of scattered electrons with respect to the directly injected hot electrons. The inset shows the dependence of (I$_{RB}$/I$_B$)$^{1/2}$ with tip bias that is found to be linear, as expected, and arises from the R-BEEM process which includes an extra $(V_T-\phi_B)^2$ factor with respect to the direct BEEM.}
\end{figure}

\begin{figure}
\includegraphics[scale=0.36]{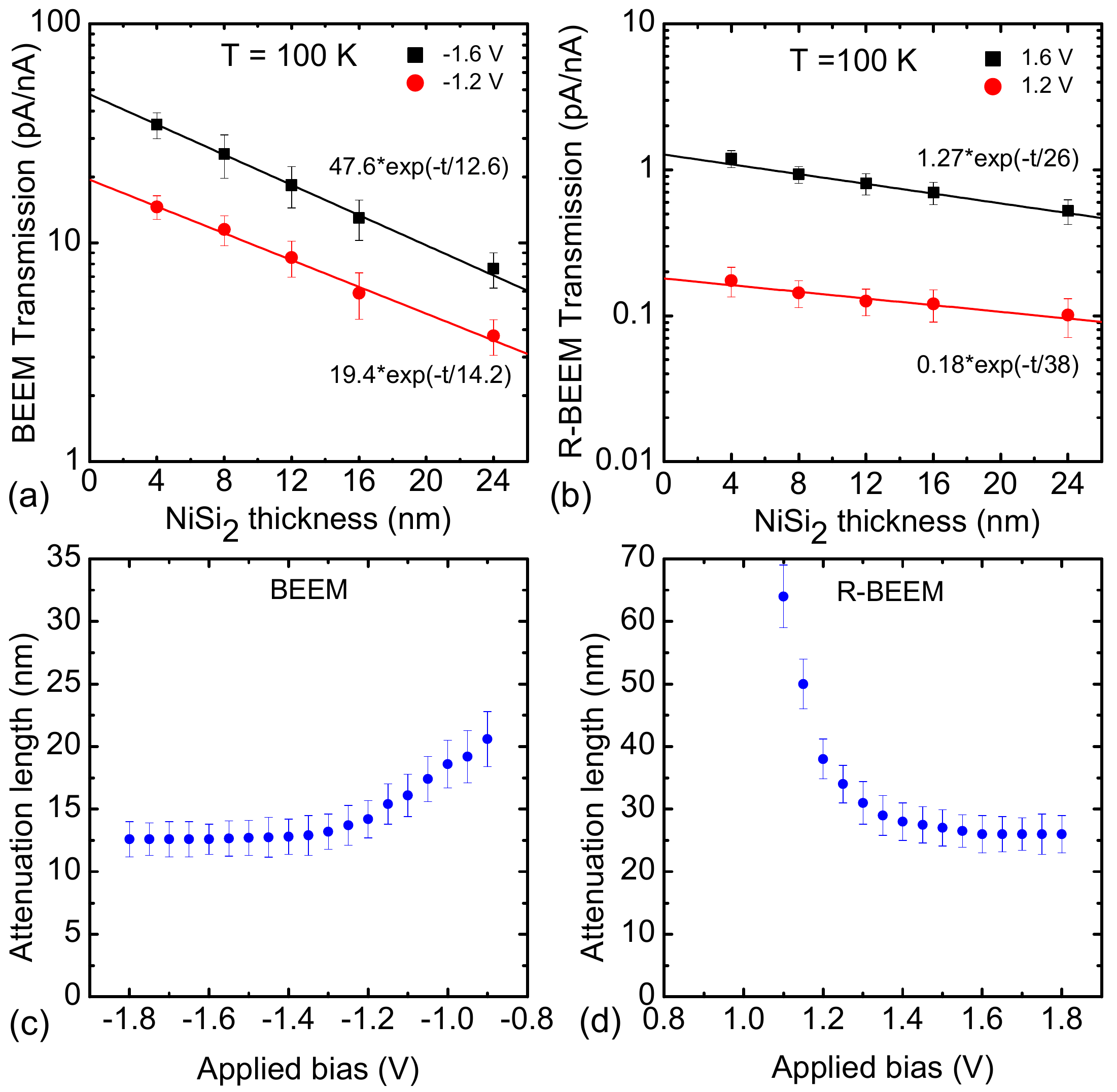}
\caption{\label{fig:TEM} (a) BEEM transmissions as function of the thickness of the NiSi$_2$ layers in the direct BEEM and (b) for the reverse BEEM. The data shown are for the same voltage in both cases. (c) The hot electron attenuation lengths as function of the applied tip bias for the direct BEEM and (d) for the reverse BEEM.}
\end{figure}

\end{document}